\documentclass[epjCONF]{svjour}

\usepackage{amsmath}
\usepackage[varg]{txfonts} % Times fonts
\usepackage[latin1]{inputenc}
\usepackage{bm}				    % for bold mathematical formulas
\usepackage{amssymb}
\usepackage{latexsym}
\usepackage{graphics}

\begin{document}

\session-title{Hot and Cold Baryonic Matter -- HCBM 2010}

\title{Zero temperature properties of mesons in a vector meson
  extended linear sigma model}

\author{P. Kov{\'a}cs\inst{1}\fnmsep\thanks{\email{pkovacs@kfki.rmki.hu}}
  \and G. Wolf\inst{1} \and F. Giacosa\inst{2} \and Denis
  Parganlija\inst{2} }

\institute{Research Institute for Particle and Nuclear Physics, H-1525
  Budapest, POB 49, Hungary \and Institute for Theoretical Physics,
  Johann Wolfgang Goethe University, Max-von-Laue-Str. 1, D--60438
  Frankfurt am Main }

\abstract{ A three flavor linear sigma model with vector and
  axial-vector mesons is discussed. Preliminary results concerning on
  the symmetry breaking pattern, the question of parameterization, as
  well as the resulting meson masses are presented.  }

\maketitle

\section{Introduction}
\label{sec:intro}

Effective field theories play a very important role in the
investigation of the strong interaction \cite{geffen_1969}, since in
the fundamental theory (QCD) lots of questions can not be answered
directly due to the complexity of the model. For instance up to now it
is still unknown how the mesons and hadrons are built up from the
basic degrees of freedom, namely from quarks and gluons. However, in
effective field theories, which possess the same global symmetries
(chiral symmetry) as QCD, the meson/hadron spectrum can be
investigated thoroughly.

The meson vacuum phenomenology can be analyzed very well in the
framework of linear sigma model \cite{levy_1967}. In this model the
global $U_L(3)\times U_R(3)$ symmetry of the massless QCD is realized
linearly. Since the $U_L(3)\times U_R(3)$ symmetry is broken due to
the axial anomaly \cite{thooft_1976} to $SU_A(3)\times U_V(3)$, a
$U_A(1)$ breaking term is introduced into the Lagrangian of the
effective model (see \cite{lenaghan_2000} and references\\
therein). The meson fields of the model are placed in $3\times 3$
matrices (nonets), which transform according to the adjoint
representation of $U_L(3)\times U_R(3)$. In the present investigation
we use an extended version of the linear sigma model, which includes
besides the usual scalar and pseudoscalar nonets a vector and an
axial-vector nonet as well. Thus we are taking into account all the
low lying mesonic degrees of freedom.

The experimental data on the majority of mesons are well established
\cite{PDG}, however there is still some open questions. For instance
the structure of the scalar mesons is still ambiguous
\cite{rischke_2010}. In this paper we present in short our calculation
in the extended linear sigma model (eLSM) concerning on the meson spectrum. A
more complete analysis will come shortly \cite{prep}.

The paper is organized as follows, in Sec.~\ref{sec:model} the model
and the symmetry breaking pattern is presented. In
Sec.~\ref{sec:masses} the tree-level masses are presented with some
remarks on the parameterization. We conclude in Sec.~\ref{sec:conc}.

\section{The model}
\label{sec:model}

Our starting point is the $U_L(3)\times U_R(3)$ symmetric linear sigma
model with vector and axial-vector degrees of freedom, and determined
by the following Lagrangian
\begin{align}
  \label{eq:lag}
  \mathcal{L} & =
  \mathrm{Tr}\left[(D^{\mu}\Phi)^{\dagger}(D^{\mu}\Phi)\right]-m_{0}^{2}
  \mathrm{Tr}(\Phi^{\dagger}\Phi)-\lambda_{1}\left[\mathrm{Tr}(\Phi^{\dagger} \Phi)\right]^{2}\nonumber\\
  & - \lambda_{2}\mathrm{Tr} \left[(\Phi^{\dagger}\Phi)^{2}\right]
  + c(\det\Phi+\det\Phi^{\dagger}) + \mathrm{Tr}\left[H(\Phi+\Phi ^{\dagger})\right] \nonumber\\
  & -
  \frac{1}{4}\mathrm{Tr}\left[(L^{\mu\nu})^{2}+(R^{\mu\nu})^{2}\right]
  + \frac{m_{1}^{2}}{2} \mathrm{Tr}
  \left[(L^{\mu})^{2}+(R^{\mu})^{2}\right]\nonumber\\
  &
  +\frac{\xi_{1}}{2}\mathrm{Tr}(\Phi^{\dagger}\Phi)\mathrm{Tr}[(L^{\mu})^{2}
  +(R^{\mu})^{2}]+\xi_{2}\mathrm{Tr}[(\Phi R^{\mu})^{2}+(L^{\mu}\Phi)^{2}]\nonumber\\
  & + 2\xi_{3}\mathrm{Tr}(\Phi R_{\mu}\Phi^{\dagger}L^{\mu}) +
  \mathcal{L}_{3} + \mathcal{L}_{4},
\end{align}

\noindent where

\begin{align}
  \Phi &= \sum_{i=0}^{8}(\sigma_i+i\pi_i^{\mu})T_i, \nonumber\\
  R^{\mu} &= \sum_{i=0}^{8}(\rho_i^{\mu}-b_i^{\mu})T_i, \nonumber\\
  L^{\mu} &= \sum_{i=0}^{8}(\rho_i^{\mu}+b_i^{\mu})T_i, \nonumber\\
  H &= \sum_{i=0}^{8}h_iT_i, \\
  D^{\mu}\Phi &= \partial^{\mu}\Phi-ig_{1}(L^{\mu}\Phi-\Phi R^{\mu})-ieA^{\mu} [T_{3},\Phi],\nonumber\\
  L^{\mu\nu} & = \partial^{\mu}L^{\nu}-ieA^{\mu}[T_{3},L^{\nu}]
  -\left\{
    \partial^{\nu}L^{\mu}-ieA^{\nu}[T_{3},L^{\mu}]\right\}, \nonumber\\
  R^{\mu\nu} & = \partial^{\mu}R^{\nu}-ieA^{\mu}[T_{3},R^{\nu}]
  -\left\{
    \partial^{\nu}R^{\mu}-ieA^{\nu}[T_{3},R^{\mu}]\right\},\nonumber
\end{align}
and $T_i\, (i=0\dots 8)$ are the generators of $U(3)$. Moreover,
$\sigma_i$ stands for the scalar, $\pi_i$ for the pseudoscalar,
$\rho^{\mu}_i$ for the vector, and $b^{\mu}_i$ for the axial-vector
mesons, while $A^{\mu}$ is the electromagnetic field and $h_i$ are the
constant external fields. $\mathcal{L}_{3}$ and $\mathcal{L}_{4}$ in
Eq.~(\ref{eq:lag}) contain three and four couplings of the different
fields, the explicit forms of which are irrelevant in our present
investigation (see e.g.~\cite{rischke_2010}).

In Eq.~(\ref{eq:lag}) there are two terms, which breaks the original
$U_L(3)\times U_R(3)$ symmetry, namely the fifth term (the determinant
term), and the sixth term (the explicit symmetry breaking term). The
first one breaks the $U_A(1)$ symmetry, while the second one breaks
the complete $U_A(3)$ if $h_0\ne 0$ and $U_V(3)\to S\negthinspace
U_V(2)\times U_V(1)$ if $h_8\ne 0$ (for details see
e.g.~\cite{lenaghan_2000}).

The meson fields ($\sigma_i,\pi_i,\rho^{\mu}_i,b^{\mu}_i$) do not have
well defined quantum numbers that can be obtained with a block
diagonal transformation of the form $f_{A} = B_{Ai}f_i,$ $f \in
(\sigma, \pi,$ $\rho^{\mu}, b^{\mu})$,
where \begin{equation}\label{eq:transf} B = \mathrm{diag} (1, \tau, 1,
  \tau, \tau, 1),\, \tau = \frac{1}{\sqrt{2}}\left( \begin{matrix} 1 &
      -i \\ 1 & i \end{matrix} \right). \end{equation} As $A$ goes
from $0$ to $8$ the components of the meson fields goes through on the
well known physical particles (for instance in case of the
pseudoscalars: $\pi_0,$ $\pi^+,$ $\pi^-,$ $\pi^0,$ $K^+,$ $K^-,$
$K^0,$ $\bar{K}^0,$ $\pi_8$), except in the $0-8$ sector, where there
is mixing between the particles (in case of the pseudoscalars this
means that only a certain linear combination of $\pi_0$ and $\pi_8$
will be mass eigenstates). For our calculations it is more suitable to
choose another base in the $0-8$ sector, which is called the non
strange\,-\,strange base, and it is given by the following linear
transformation,
\begin{eqnarray}
  f_N &=\sqrt{2/3}f_0+\sqrt{1/3}f_8,\nonumber \\
  f_S &=\sqrt{1/3}f_0 - \sqrt{2/3}f_8,\label{eq:nsbase}
\end{eqnarray}
where $f \in (\sigma, \pi,$ $\rho^{\mu}, b^{\mu})$. To see more
explicitly the structure of the $\Phi$, $L^{\mu}$, $R^{\mu}$ fields,
we give their matrix form (see \cite{proc1}),

\begin{eqnarray}
  \Phi =\frac{1}{\sqrt{2}}\left( 
    \begin{array}{ccc}
      \frac{(\sigma_N+a_{0}^{0})+i(\pi_{N}+\pi ^{0})}{\sqrt{2}} & 
      a_{0}^{+}+i\pi ^{+} & K_{S}^{+}+iK^{+} \\ 
      a_{0}^{-}+i\pi ^{-} & \frac{(\sigma_N-a_{0}^{0})+i(\pi_{N}-\pi ^{0})}{
        \sqrt{2}} & K_{S}^{0}+iK^{0} \\ 
      K_{S}^{-}+iK^{-} & {\bar{K}_{S}^{0}}+i{\bar{K}^{0}} & \sigma _{S}+i\pi_{S}
    \end{array}
  \right),    \nonumber\\
  L^{\mu } =\frac{1}{\sqrt{2}}\left( 
    \begin{array}{ccc}
      \frac{\rho_{N}+\rho ^{0}}{\sqrt{2}}+\frac{a_{1N}+a_{1}^{0}}{\sqrt{2}} & 
      \rho ^{+}+a_{1}^{+} & K^{\star +}+K_{1}^{+} \\ 
      \rho ^{-}+a_{1}^{-} & \frac{\rho_{N}-\rho^{0}}{\sqrt{2}}+\frac{
        a_{1N}-a_{1}^{0}}{\sqrt{2}} & K^{\star 0}+K_{1}^{0} \\ 
      K^{\star -}+K_{1}^{-} & {\bar{K}}^{\star 0}+{\bar{K}}_{1}^{0} & \rho_{S}+a_{1S}
    \end{array}
  \right)^{\mu}, \label{eq:matrix} \\
  R^{\mu } =\frac{1}{\sqrt{2}}\left( 
    \begin{array}{ccc}
      \frac{\rho_{N}+\rho ^{0}}{\sqrt{2}}-\frac{a_{1N}+a_{1}^{0}}{\sqrt{2}} & 
      \rho ^{+}-a_{1}^{+} & K^{\star +}-K_{1}^{+} \\ 
      \rho ^{-}-a_{1}^{-} & \frac{\rho_{N}-\rho ^{0}}{\sqrt{2}}-\frac{
        a_{1N}-a_{1}^{0}}{\sqrt{2}} & K^{\star 0}-K_{1}^{0} \\ 
      K^{\star -}-K_{1}^{-} & {\bar{K}}^{\star 0}-{\bar{K}}_{1}^{0} & \rho_{S}-a_{1S}
    \end{array}
  \right)^{\mu}.  \nonumber
\end{eqnarray}
The experimentally observed mesons can be assigned to the above fields
as follows, $\pi^{\pm},\pi^0$ and $K^{\pm},K^0,\bar{K}^0$ corresponds
to the well-known pion $(\pi(138))$ and kaon $(K(496))$,
respectively. The $\pi_0, \pi_8$ fields are mixture of the $\eta(548)$
and $\eta^{\prime}(958)$ particles. In the scalar sector the
assignment is not so obvious, since there are more than one candidate
for every scalar fields. In accordance with \cite{rischke_2010}, where
the scalar states were found to be above $1$~GeV, we can assign
$K^{\star\pm}_S,K_S^{\star 0},\bar{K}_S^{\star 0}$ to the
$K_S^{\star}(1430)$, while $a_0^{\pm}, a_0^0$ possibly to the
$a_0(1450)$, respectively. In this sector the mixture of $\sigma_0$
and $\sigma_8$ can form the $f_0(1370)$ and $f_0(1710)$
particles. Since this sector is the most uncertain, we would like to
use as few of them as it is possible for the parameterization (see
Sec.~\ref{sec:masses}), and treat them instead as predictions. The
$\rho^{\mu\,\pm},\rho^{\mu\,0}$ and $K^{\star\pm},K^{\star
  0},\bar{K}^{\star 0}$ fields represent the $\rho(770)$ and
$K^{\star}(892)$ vector mesons, respectively. The remaining two vector
meson fields $\rho^{\mu}_0$ and $\rho^{\mu}_8$ are the mixture of the
$\Phi(1020)$ and $\omega(782)$ particles. Finally, the axial-vector
meson fields $a_1^{\mu\,\pm}, a_1^{\mu\,0}$ and $K_{1}^{\mu\,\pm},
K_1^{\mu\,0}, \bar{K}_1^{\mu\, 0}$ correspond to the $a_1(1260)$ and
$K_1(1270)$, respectively, while $a_{1, 0}^{\mu}$ and $a_{1, 8}^{\mu}$
are mixture of $f_1(1285)$ and $f_1(1420)$.

\subsection{Symmetry breaking}
\label{sec:sym_break}

In this model the chiral symmetry is broken explicitly (the sixth term
of Eq.~(\ref{eq:lag})) as well as spontaneously. In case of
spontaneous symmetry breaking the effective potential has its minimum
at a non-vanishing value, which corresponds to a non-zero expectation
value for some of the fields. Since the vacuum has zero quantum
numbers the possible fields are the $\sigma_0$ and $\sigma_8$ scalar
fields \cite{lenaghan_2000}. Let us denote the expectation values for
$\sigma_0$ and $\sigma_8$ as $\Phi_0$ and $\Phi_8$,
respectively. However, as in case of the fields, it is more convenient
to use the non strange\,-\,strange base (see Eq.~(\ref{eq:nsbase})).

According to the usual process, we shift $\sigma_N$ and $\sigma_S$ by
their vacuum expectation values $\Phi_N$ and $\Phi_s$ and substitute
into the Lagrangian. This will result in a technical difficulty,
namely that mixing terms appear among certain fields in the
Lagrangian.

\section{Tree-level masses}
\label{sec:masses}

In order to calculate the tree-level masses after the introduction of
the shifts, all the quadratic terms must be considered, which can be
written as,
\begin{align}
  \mathcal{L}^{\mathrm{quad}} &= -\frac{1}{2} \sigma_A \left(
    \delta_{AB} \partial^2 + (m_\sigma^2)_{AB} \right)\sigma_B
  \nonumber \\
  &- \frac{1}{2} \pi_A \left( \delta_{AB} \partial^2 + (m_\pi^2)_{AB}
  \right) \pi_B \nonumber \\
  & - \frac{1}{2}\rho_{A\mu}\left((-g^{\mu\nu}\partial^2
    + \partial^\mu \partial^\nu)\delta_{AB}-g^{\mu\nu}(m_\rho^2)_{AB}\right)\rho_{B\nu}
  \nonumber \\
  & - \frac{1}{2}b_{A\mu}\left((-g^{\mu\nu}\partial^2
    + \partial^\mu \partial^\nu)\delta_{AB}-g^{\mu\nu}(m_b^2)_{AB}\right)b_{B\nu}
  \label{eq:quad}\\
  & - \frac{1}{2} \rho_{A\mu}\left(ig_1 f_{ABC} \Phi_C \partial^{\mu}
  \right) \sigma_B - \frac{1}{2} \sigma_A\left(ig_1
    f_{ABC}\Phi_C\partial^{\nu}\right)\rho_{B\nu}\nonumber\\
  & + \frac{1}{2} b_{A\mu}\left(g_1 d_{ABC} \Phi_C \partial^{\mu}
  \right) \pi_B - \frac{1}{2} \pi_A\left(g_1
    d_{ABC}\Phi_C\partial^{\nu}\right)b_{B\nu},\nonumber
\end{align}

\noindent where

\begin{align}
  (m_\sigma^2)_{AB} & = m_0^2\delta_{AB} - 6 G_{ABC}\Phi_C + 4
  F_{ABCD} \Phi_C\Phi_D\\
  (m_\pi^2)_{AB} & = m_0^2\delta_{AB} + 6 G_{ABC}\Phi_C + 4
  H_{AB,CD} \Phi_C\Phi_D\\
  (m_\rho^2)_{AB} & = m_1^2\delta_{AB} + g_1^2
  f_{ACM}f_{BDM}\Phi_C\Phi_D + 2 J_{AB,CD} \Phi_C\Phi_D\\
  (m_b^2)_{AB} & = m_1^2\delta_{AB} + g_1^2 d_{ACM}d_{BDM}\Phi_C\Phi_D
  + 2 J^{\prime}_{AB,CD} \Phi_C\Phi_D.
\end{align}

Here $\Phi_A$ denotes the vector $(\Phi_N,0,0,0,0,0,0,0,\Phi_S)$,
while $f_{ABC}$ and $d_{ABC}$ are the antisymmetric and symmetric
group structure constants transformed by (\ref{eq:transf}),
viz. $f_{ABC}=f_{abc} B^{-1}_{aA} B^{-1}_{bB} B^{-1}_{cC}$, and
$d_{ABC}=d_{abc} B^{-1}_{aA} B^{-1}_{bB} B^{-1}_{cC}$. The $G_{ABC}$,
$F_{ABCD}$, $H_{AB,CD}$, $J_{AB,CD}$, and $J^{\prime}_{AB,CD}$
coefficient tensors contain only the group structure constants and the
coupling constants of the Lagrangian. The first two coefficient
tensors $G$ and $F$ are totally symmetric, while $H$, $J$, and
$J^{\prime}$ are symmetric in the first two and in the second two
indices.

The last four terms of Eq.~(\ref{eq:quad}) are mixing terms between
different types of mesons. There are two-two terms for the
vector-scalar, and for the axial-vector-pseudoscalar mixing. Using the
explicit forms of $f_{ABC}$ and $d_{ABC}$ the following mixings are
present,
\begin{align}
  \pi_{N}-a_{1N}^{\mu} &: -g_{1}\phi_{N}a_{1N}^{\mu }\partial_{\mu}\pi_{N},\nonumber \\
  \pi-a_{1}^{\mu} &: -g_{1}\phi_{N}({a_{1}^\mu}^{+}\partial_{\mu
  }\pi^- +
  {a_{1}^\mu}^{0}\partial_{\mu }\pi^0) + \mathrm{h.c.},\nonumber\\
  \pi_{S}-a_{1S}^{\mu} &: -\sqrt{2}g_{1} \phi_{S}
  a_{1S}^{\mu}\partial_{\mu }\pi_{S},\\
  K_{S}-K^{\star}_{\mu} &: \frac{ig_{1}}{2}(\sqrt{2} \phi_{S} -
  \phi_{N}) (\bar{K}^{\star 0}_{\mu}\partial^{\mu} K_{S}^{0} +
  K^{\star
    -}_{\mu} \partial^{\mu}K_{S}^{+}) + \mathrm{h.c.},\nonumber\\
  K-K_1^{\mu} &:-\frac{g_{1}}{2}(\phi _{N} + \sqrt{2}\phi _{S})
  (K_{1}^{\mu 0}\partial_{\mu } \bar{K}^{0} + K_{1}^{\mu
    +} \partial_{\mu } K^{-}) + \mathrm{h.c.}.\nonumber
\end{align}
These mixings can be resolved by appropriate transformations for the
${K^{\star}}^{\mu}$ vector and the $a_1^{\mu}$, $a_{1S/N}^{\mu}$, and
$K_1^{\mu}$ axial-vector meson fields. The necessary transformations
are the following,
\begin{align}
  a_{1N/S}^{\mu} &\longrightarrow a_{1N/S}^{\mu} +
  w_{a_{1N/S}} \partial^{\mu}\pi_{N/S},\nonumber\\
  {a_1^{\mu}}^{\pm,0}&\longrightarrow {a_1^{\mu}}^{\pm,0} +
  w_{a_1} \partial^{\mu} \pi^{\pm,0},\nonumber\\
  {K_1^{\mu}}^{\pm,0}&\longrightarrow {K_1^{\mu}}^{\pm,0} +
  w_{K_1} \partial^{\mu} K^{\pm,0},\nonumber\\
  {\bar{K}_{1}^{\mu 0}}&\longrightarrow {\bar{K}_1^{\mu 0}} +
  w_{K_1} \partial^{\mu} \bar{K}^{0},\label{eq:trans_vec}\\
  {K^{\star\mu}}^{+}&\longrightarrow {K^{\star\mu}}^{+} +
  w_{K^{\star}} \partial^{\mu} K_S^{+},\nonumber\\
  {K^{\star\mu}}^{-}&\longrightarrow {K^{\star\mu}}^{-} +
  w^{\star}_{K^{\star}} \partial^{\mu} K_S^{-},\nonumber\\
  {K^{\star\mu}}^{0}&\longrightarrow {K^{\star\mu}}^{0} +
  w_{K^{\star}} \partial^{\mu} K_S^{0},\nonumber\\
  \bar{K}^{\star\mu 0}&\longrightarrow \bar{K}^{\star\mu 0} +
  w^{\star}_{K^{\star}} \partial^{\mu} \bar{K}_S^{0}.\nonumber
\end{align}
After transforming the fields with (\ref{eq:trans_vec}) in
(\ref{eq:quad}), the $w_{A}$ coefficients can be determined by
requiring the disappearance of the mixed terms. It is important to
note that this transformation leads to the appearance of
multiplicative factors in front of the kinetic terms of the $\pi$,
$\pi_N$, $\pi_S$, $K$, and $K_S$ fields, in other words after the
transformations they are not canonically normalized anymore. The
multiplicative factors are denoted as $Z_{\pi}$, $Z_{\pi_N}$,
$Z_{\pi_S}$, $Z_K$ and $Z_{K_S}$. These factors are similar that of
the wave function renormalization constants, however, they can take
larger values than $1$ \cite{rischke_2010} to the contrary of the
usual wave function renormalization constant (see
e.g. \cite{peskin95}). Thus in order to get the canonical scalar
propagator form, these fields must be renormalized. After a
straightforward but lengthy calculation, the coefficients are found to
be,
\begin{align}
  w_{a_{1N}}=w_{a_{1}}= \frac{g_{1}\phi _{N}}{m_{a_{1}}^{2}} ,\\
  w_{a_{1S}}=\frac{\sqrt{2}g_{1}\phi _{S}}{m_{a_{1S}}^{2}} , \\
  w_{K^{\star }}=\frac{ig_{1}(\phi _{N} - \sqrt{2}\phi
    _{S})}{2m_{K^{\star }}^{2}} , \\
  w_{K_{1}}= \frac{g_{1}(\phi _{N} + \sqrt{2}\phi
    _{S})}{2m_{K_{1}}^{2}},
\end{align}
while the renormalization factors are,
\begin{align}
  Z_{\pi } &\equiv Z_{\pi _{N}} =\frac{m_{a_{1}}}{\sqrt{m_{a_{1}}^{2}
      - g_{1}^{2}\phi_{N}^{2}}}, \\
  Z_{\pi_{S}} & =\frac{m_{a_{1S}}}{\sqrt{m_{a_{1S}}^{2} -
      2g_{1}^{2}\phi_{S}^{2}}}, \\
  Z_{K} & =\frac{2m_{K_{1}}}{\sqrt{4m_{K_{1}}^{2} - g_{1}^{2}(\phi_{N}
      + \sqrt{2}\phi_{S})^{2}}},   \\
  Z_{K_{S}} & =\frac{2m_{K_{\star}}}{\sqrt{4m_{K_{\star}}^{2} -
      g_{1}^{2}(\phi_{N} - \sqrt{2}\phi _{S})^{2}}}.
\end{align}
Using the following notations, $\Lambda_N \equiv
\lambda_1+\lambda_2/2$, $\Lambda^{\prime}_N \equiv
\lambda_1+3\lambda_2/2$, and $\Lambda_S \equiv \lambda_1+\lambda_2$,
the tree-level pseudoscalar masses are obtained as
\begin{align}
  m_{\pi}^2 &= Z_{\pi}^2\left[m_0^2 + \Lambda_N\Phi_N^2 +
    \lambda_1\Phi_S^2 -
    \frac{c}{\sqrt{2}} \Phi_S \right],\\
  m_{K}^2 &= Z_K^2\left[m_0^2 + \Lambda_N\Phi_N^2 -
    \frac{\lambda_2}{\sqrt{2}}
    \Phi_N\Phi_S + \Lambda_S\Phi_S^2 - \frac{c}{2} \Phi_N \right],\\
  m_{\pi_N}^2 &= Z_{\pi}^2\left[m_0^2 + \Lambda_N\Phi_N^2 +
    \lambda_1\Phi_S^2 +
    \frac{c}{\sqrt{2}} \Phi_S \right],\\
  m_{\pi_S}^2 &= Z_{\pi_S}^2\left[m_0^2 + \lambda_1\Phi_N^2 +
    \Lambda_s\Phi_S^2 \right],\\
  m_{\pi_{NS}}^2 &= Z_{\pi}Z_{\pi_S}\frac{c}{\sqrt{2}}\Phi_N,
\end{align}
while the scalar masses are,
\begin{align}
  m_{a_0}^2 &= m_0^2 + \Lambda^{\prime}_N\Phi_N^2 + \lambda_1\Phi_S^2
  +
  \frac{c}{\sqrt{2}} \Phi_S,\\
  m_{K_S}^2 &= Z_{K_S}^2\left[m_0^2 + \Lambda_N\Phi_N^2 +
    \frac{\lambda_2}{\sqrt{2}} \Phi_N\Phi_S + \Lambda_S\Phi_S^2 +
    \frac{c}{2} \Phi_N \right],
\end{align}
\begin{align}
  m_{\sigma_N}^2 &= m_0^2 + 3\Lambda_N\Phi_N^2 + \lambda_1\Phi_S^2 -
  \frac{c}{\sqrt{2}} \Phi_S,\\
  m_{\sigma_S}^2 &= m_0^2 + \lambda_1\Phi_N^2 +
  3\Lambda_s\Phi_S^2,\\
  m_{\sigma_{NS}}^2 &=
  2\lambda_1\Phi_N\Phi_S-\frac{c}{\sqrt{2}}\Phi_N,
\end{align}
where the $m_{\pi_{NS}}^2$, and $m_{\sigma_{NS}}^2$ are mixing terms
in the non-strange-strange sector. These mixings can be removed by
orthogonal transformations, and the resulting mass eigenstates are
found to be,
\begin{align}
  m^{2}_{{f_{0}^{H}}/{f_{0}^{L}}} &= \frac{1}{2}\left[ m_{\sigma_N}^2
    + m_{\sigma_S}^2 \pm \sqrt{(m_{\sigma_N}^2-m _ {\sigma_S}^2)^2 + 4
      m_{\sigma_{NS}}^2 } \right],\\
  m^{2}_{\eta^{\prime}/\eta} &= \frac{1}{2}\left[ m_{\pi_N}^2 +
    m_{\pi_S}^2 \pm \sqrt{(m_{\pi_N}^2-m _ {\pi_S}^2)^2 + 4
      m_{\pi_{NS}}^2 } \right].
\end{align}
Using the notations $\Xi_N \equiv (g_1^2/2+\xi_1+\xi_2/2)/2$, $\Xi_S
\equiv (g_1^2 + \xi_1 + \xi_2)/2$, the vector masses are found to be,
\begin{align}
  m_{\rho}^2 &= m_1^2 + \frac{1}{2} (\xi_1+\xi_2+\xi_3) \Phi_N^2 +
  \frac{\xi_1}{2} \Phi_S^2,\\
  m_{K^{\star}}^2 &= m_1^2 + \Xi_N \Phi_N^2
  +\frac{1}{\sqrt{2}} \Phi_N\Phi_S(\xi_3 - g_1^2) + \Xi_S \Phi_S^2,\\
  m_{\omega}^2 &= m_{\rho}^2,\\
  m_{\Phi}^2 &= m_1^2 + \frac{\xi_1}{2} \Phi_N^2 +
  \left(\frac{\xi_1}{2}+\xi_2+\xi_3\right) \Phi_S^2,
\end{align}
and finally the axial-vector meson masses are given by,
\begin{align}
  m_{a_1}^2 &= m_1^2 + \frac{1}{2} (2g_1^2+\xi_1+\xi_2-\xi_3) \Phi_N^2
  +
  \frac{\xi_1}{2} \Phi_S^2,\\
  m_{K_1}^2 &= m_1^2 + \Xi_N \Phi_N^2
  -\frac{1}{\sqrt{2}} \Phi_N\Phi_S(\xi_3 - g_1^2) + \Xi_S \Phi_S^2,\\
  m_{f_1^{L}}^2 &= m_{a_1}^2,\\
  m_{f_1^H}^2 &= m_1^2 + \frac{\xi_1}{2} \Phi_N^2 + \left(2g_1^2 +
    \frac{\xi_1}{2}+\xi_2-\xi_3\right) \Phi_S^2.
\end{align}
It is worth to note that in case of vectors and axial-vectors there
are no mixing terms in the non strange-strange sector.
\subsection{Parameterization}
\label{sec:param}
In order to calculate the tree-level masses in physical units, the
unknown parameters of the model must be determined. There are eleven
unknown parameters, namely $m_0^2$, $m_1^2$, $c$, $g_1$, $\lambda_1$,
$\lambda_2$, $\xi_1$, $\xi_2$, $\xi_3$ and the two condensates $\Phi_N$,
$\Phi_S$. Since, there are 14 different masses in our model, and all of
them are expressed with these parameters, one can choose an
appropriate set (experimentally well established masses), and treat
them as a system of equations for the parameters. The system of
equations can be solved with multi-parametric minimalization. This work
is still ongoing, however some preliminary results can be found in
\cite{proc2}.

\section{Conclusion}

We have presented a three flavor linear sigma model with vector and
axial-vector degrees of freedom. Implementing the spontaneous symmetry
breaking in the model yields not only the known $\pi_{N}$-$a_{1N}$ and
$\pi^{\pm,0}$-$a_{1}^{\pm,0}$ mixings \cite{rischke_2010} but also the
$\pi_{S}$-$a_{1S}$, $K_{S}$-$K^{\star }$ and $K$-$K_{1}$ mixings as well. By
using the transformations Eq.~(\ref{eq:trans_vec}), and subsequently
bringing the $\pi_{N,S}$, $\pi$, $K_{S}$ and $K$ derivatives to the
canonical form, the non-diagonal terms in the Lagrangian can be
removed, which leads to the introduction of the pion, kaon, scalar kaon
renormalization coefficients. $Z_{\pi}, Z_{\pi_N}, Z_{\pi_S}, Z_{K},
Z_{K_S}$. The tree-level masses than can be expressed with the eleven
unknown parameters of the model, which can be determined by using the
experimentally well known particle masses \cite{PDG} and multi-parametric
minimalization. Detailed analysis of the different parameterizations and
calculations of the decay widths of the resonances in the Lagrangian
(\ref{eq:lag}) will be presented in a separate work \cite{prep}.

\label{sec:conc}

% \begin{table}
%\caption{Please write your table caption here.}
%\begin{tabular}{lll}
%\hline\noalign{\smallskip}
%\end{tabular}
%\end{table}
%

\end{document}